# Rapid Identification of X-ray Diffraction Spectra Based on Very Limited Data by Interpretable Convolutional Neural Networks

Hong Wang[1#], Yunchao Xie[1#], Dawei Li[1], Heng Deng[1], Yunxin Zhao[2], Ming Xin[1], and Jian Lin[1,2,3*]

[1]Department of Mechanical and Aerospace Engineering
[2]Department of Electrical Engineering and Computer Science
[3]Department of Physics and Astronomy
University of Missouri, Columbia, Missouri 65211, USA
*E-mail: LinJian@missouri.edu (J. L.)
#Authors contributed equally to this work.

**Abstract**

Large volumes of data from material characterizations call for rapid and automatic data analysis to accelerate materials discovery. Herein, we report a convolutional neural network (CNN) that was trained based on theoretic data and very limited experimental data for fast identification of experimental X-ray diffraction (XRD) spectra of metal-organic frameworks (MOFs). To augment the data for training the model, noise was extracted from experimental spectra and shuffled, then merged with the main peaks that were extracted from theoretical spectra to synthesize new spectra. For the first time, one-to-one material identification was achieved. The optimized model showed the highest identification accuracy of 96.7% for the Top 5 ranking among a dataset of 1012 MOFs. Neighborhood components analysis (NCA) on the experimental XRD spectra shows that the spectra from the same material are clustered in groups in the NCA map. Analysis on the class activation maps of the last CNN layer further discloses the mechanism by which the CNN model successfully identifies individual MOFs from the XRD spectra. This CNN model trained by the data-augmentation technique would not only open numerous potential applications for identifying XRD spectra for different materials, but also pave avenues to autonomously analyze data by other characterization tools such as FTIR, Raman, and NMR.

**Introduction**

High-throughput synthesis techniques have shown great potential in accelerating material innovation.[1] Large volumes of characterization data including X-ray diffraction (XRD), Raman, nuclear magnetic resonance (NMR), and Fourier Transform Infrared (FTIR) spectra are collected during or after the synthesis. Among them, XRD is a powerful technique to characterize crystallographic structures, grain size, and molecular structures.[2] Typically, experimental XRD patterns are analyzed *via* comparing descriptors such as peak positions, intensities, full widths at half maximum (FWHM) against a known database such as Crystallography Open Database and Inorganic Crystal Structure Database, allowing scientists to identify the compounds of interest and to map phase diagrams of combinatorial materials. However, the tedious and time-consuming procedure due to the manual analysis at a relatively low speed severely hinders the fast decision making.[2, 3] To fully exploit the characterization tools, it is becoming urgent to develop new data assessment tools with automation and recommendation functions, especially with the emerging of self-driven laboratories enabled by robots.[4-6] Despite recent progress, it has been and continues to be a grand challenge.

Recently, machine learning (ML) models have shown great potentials in managing the large volumes of characterization data for rapidly and automatically identifying composition-phase maps as well as constructing composition-structure-property relationships, thereby speeding up the materials discovery.[1, 7-15] For instance, Park *et al*. demonstrated well-trained convolutional neural networks (CNNs) which exhibited satisfactory accuracy in classifying XRD spectra based on theoretical database.[16] Oviedo and his colleagues proposed a machine learning approach to predict crystallographic dimensionality and space groups from a limited number of thin-film XRD spectra.[2] Angelo's research group developed a robust CNN model to classify crystal structures and also unfolded the internal behavior of the classification model through visualization.[14] Miller's research group implemented a CNN to determine crystallography trained on imaging and diffraction data.[15] However, these approaches were applied to identify several classes or crystal systems into which target materials are grouped. One-by-one identification of individual spectrum from millions of spectrum databases is still challenging. Another big challenge for developing the machine learning enabled methodology is the lack of experimental data for training the models. Although a technique of Gaussian mixture was employed to augment the theoretic data,[17] it may not fully reflect the real experiments when distinguished features arise from the experiments. It is envisioned that directly incorporating experimental data into theoretic data is a better approach. Finally, the deep learning models like CNN are usually treated as a "black-box". Interpreting the underlying mechanism of such as black-box for decision-making or obtaining the final desired results is still an open problem. Therefore, developing a procedure that can better interpret the deep learning models when they are applied to material research has recently seen a resurgence.

In this paper, we propose a CNN model that was trained for rapid one-to-one identification of experimental XRD spectra of metal-organic frameworks (MOFs). To increase robustness of the CNN model, noise was extracted from the experimental spectra to augment the theoretic spectra for training. In the cases of very noisy experimental spectra, the fast Fourier transform (FFT) was applied to reduce the noise before they were input into the CNN for improving the prediction accuracy. The optimized CNN model showed the highest identification accuracy of 96.7% for the Top 5 ranking among a dataset of 1012 MOFs. Data dimension reduction analysis on the experimental XRD spectra by the neighborhood components analysis (NCA) shows that the spectra from the same MOF are clustered in individual groups in the NCA map, while the XRD spectra from different MOFs but with very similar characteristics may have overlapping. Further analysis on the class activation maps (CAMs) of the fourteenth layer of the CNN model shows that the grouped spectra that are highly distinguishable in the NCA map exhibit very different activation characteristics. This observation can well explain why the CNN can identify individual spectra from the library.

The novelty of this work can be summarized as follows. First, to the best of our knowledge, this is the first demonstration that a CNN enables one-by-one identification of XRD spectrum for individual materials. The previous reported machine learning algorithms only classify several classes or crystal systems into which target materials are grouped. Second, the model was trained by theoretical spectra combined with very limited experimental data. Third, the noise-based data augmentation technique is very easy and straightforward to implement, but results in very effective outcomes. Fourth, the trained CNN model can successfully and robustly perform one-by-one classification with the help of noise filtering procedure even though the experimental XRD spectra exhibit peak shift, scaling in peak intensities, or FWHM broadening compared to the theoretic spectra. Lastly but not the least, the study on the CAMs of the convolutional layers discloses the mechanism of how the CNN model makes the decision, thus shedding new light on the interpretable deep learning for materials characterization data analysis. Consequently, the proposed solution is of great interest and appears to be very promising, not only because of the applications of XRD to characterize different types of materials, but also the possible extension to spectra collected by other characterization techniques including Raman, NMR, and FTIR.

**Results and Discussion**

The flowchart showing the procedure for rapid identification of XRD spectra enabled by the CNN is illustrated in Fig. 1a. First, theoretical CIF files of MOFs downloaded from Cambridge

Crystallographic Data Centre (CCDC)[18] were converted into theoretical XRD spectra. Experimental XRD spectra were collected from as-synthesized MOFs powder in Bruker D8 Advance. Detailed synthesis and characterization can be found in Experimental Section. Then, the noise was extracted from experimental spectra and shuffled, then merged with the main peaks that were extracted from theoretical spectra to obtain new synthesized spectra. By this data augmentation method, sufficient training datasets were realized. The experimental XRD spectra that were used as the testing datasets were filtered to reduce the noise level if needed. The detailed procedure of data augmentation and noise reduction is described in the following paragraph. A CNN was built from scratch based on Lenet5 and VGG16.[19, 20] Its architecture is shown in Fig. 1b with hyperparameters summarized in Table S1. Basically, the CNN consists of one input layer, four convolutional layers, three fully-connected layers, and one output layer. The first layer inputs synthesized XRD spectra with 2θ ranging from 5 degrees to 50 degrees. Then the data is fed subsequently into four convolution layers with kernel filters followed by a max pooling layer. The kernels for these convolutional layers are six, sixteen, thirty-two, and sixty-four filters with a size of 5×1 and a stride of one. The max pooling layer has two filters with a stride of two and a dropout with a dropout rate of 0.2.[21-23] All convolutional layers were activated by the function of the rectified linear unit (Relu). After one flatten layer, the data was fed to three dense layers with filters of 120, 84, and 1012, respectively. The Adam optimizer was implemented to minimize the categorical cross-entropy loss function.[24, 25]

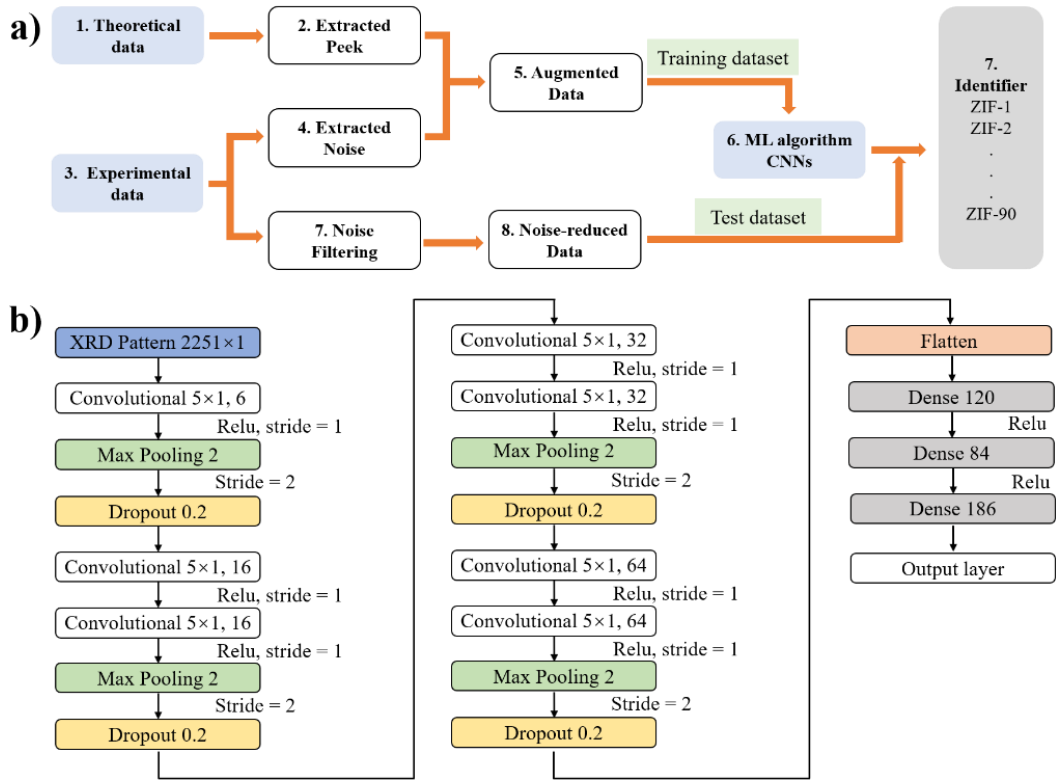

**Figure 1**. (a) Flowchart showing the process of XRD spectrum identification. (b) Architecture of proposed convolutional neural network.

Fig. 2 exhibits the detailed procedure of preprocessing and augmenting theoretic spectra, reducing the noise level of experimental spectra, and training the CNN model. The spectra used for training were synthesized by merging extracted peaks from the theoretic spectra and baseline noise from the experimental spectra. The main peaks were extracted from a theoretic spectrum containing the largest 400 points. The noise was collected from the baseline of raw experimental spectra after the main peaks were removed. Then these two components were concatenated to form a new spectrum. Since the noise can be randomly sampled and shuffled, the spectra can be largely augmented for training the CNN model. These training spectra were synthesized from a library of a total of 832 theoretical spectra. Next, FFT and inverse FFT (iFFT) were applied to smooth the experimental spectra for the purpose of reducing their noise level. FFT is a popular algorithm that converts the signal in an original domain to a representation in a frequency domain.[26] After the raw XRD spectra were converted to FFT spectra in the frequency range of 0 ~ 200 Hz, the data beyond the frequency of 200 is the white noise. Here, we use the mean peak value of the white noise as a criterion to determine whether iFFT should be applied to reduce the noise of experimental XRD spectra. If the mean value is larger than 1, it means that the noise has high intensity and the iFFT is applied to first 200 Hz. Otherwise, the original data can be directly used as testing data. A total of 24 experimental spectra collected from eight types of MOFs after noise filtering were used as the testing datasets. Compared to previous reports on the procedure of preprocessing XRD spectra, no much human intervention steps such as background removal, smoothing and interpolation, region exclusion, and peak shifting were involved,[3, 27] thereby significantly improving autonomy of the model for data analysis.

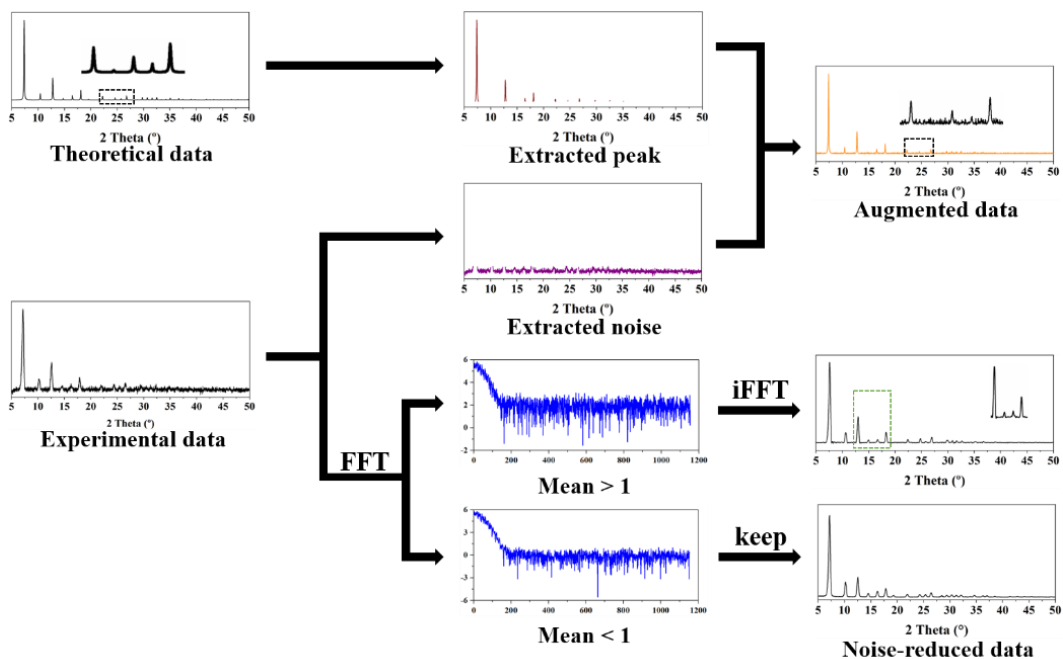

**Figure 2**. Flowchart showing process of augmenting theoretical spectra and filtering noise of experimental spectra.

After the data preprocessing, the synthesized spectra and noise-reduced experimental spectra were used as the training and test datasets, respectively, to train and test various supervised ML algorithms for evaluating one-by-one identification performance. We first tested five classical ML algorithms such as Naïve Bayes (NB), *k*-Nearest Neighbors (KNN), Logistic Regression (LR), Random Forest (RF), and Support Vector Machine (SVM) and their classification accuracies are shown in Fig. 3a and summarized in Table S2. Here, we define Top 1 to Top 5 as the ranking positions of identification results of the testing spectra among the library consisting of total 823 MOFs (Table S3). For example, Top 1 means that a ML algorithm can successfully rank a MOF sample at the first position. Different from previous study which maps the spectra into 7 crystal systems or 230 space groups among thousands of materials,[2, 16] it is much more challenging to reach the goal of one-by-one material classification. One-by-one classification mission is similar to the large-scale image-classification challenge attempted by the ImageNet, which classifies high-resolution images into 1,000 different categories. It can be seen that all five classical ML algorithms exhibit < 50% identification accuracy for Top 1-to-Top 5 rankings. In comparison, the best result of CNN performed much better with 56.7%, 76.7%, 90%, and 93.3% accuracy for Top 1, Top 2, Top 3, and Top 4 ranking, respectively (Table S4), and reached 96.7 % accuracy for Top 5 ranking. It demonstrates that high-level hidden and meaningful features learned by the CNN help identify XRD spectra with a much higher accuracy than the classical ML algorithms.[28] In addition, the classification accuracy of Top 5 over the classical ML algorithms and CNN model under different theoretical data is also investigated and shown in Fig. 3b. It is obvious that the classification accuracy of the classical ML algorithms decreases sharply when the data size increases, whereas our CNN model is robust enough to deliver predictive accuracy of 96.7% even the theoretical spectra in the library increases from 189 to 1012 (Fig. 3b). It can be expected that as the library size further increases, the prediction accuracy would be well maintained.

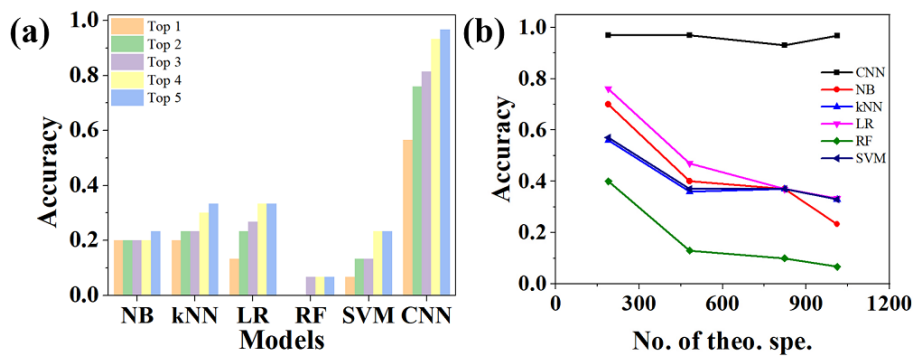

**Figure 3**. (a) Comparison of various ML models for identifying XRD spectra among the library with 1012 MOFs. NB: Navie Bayes, kNN: k-Nearest Neighbors, LOG: Logistic Regression, RF: Random Forest, SVM: Support Vector Machine, CNN: Convolutional Neural Network. (b) The Top 5 accuracy of various ML models trained with different number of theoretical spectra.

It is well-known that the size of the datasets may affect the performance of a ML algorithm.[29] Thus, we investigated the effect of the library size of the MOFs on the prediction accuracy. First, the ratio of augmented training data to corresponding theoretical data also affects the prediction accuracy. As it increases, one can identify that the accuracy quickly reaches the highest number (96.7%), while it gradually drops to 79.4 % when the ratio of augmentation is further increased to 300 (Fig. S1a). Other than the dataset size, the number of training epochs also affects the performance (Fig. S1b). With the increase of the epoch number, the identification accuracy first increases to 96.7% from 75%, then decreases to 75% again if it further increases. At the epoch number of 40, the accuracy reaches the highest value of 96.7%. Thus, the early stopping was set to 40 epochs. The identification accuracy of the convolutional layer number also exhibits the similar trend (Fig. S1c), i.e., first increases to 96.7% with the increase of the layer number to 7, then decreases to 77.8% as the number of the layers increases to 9. Hence, the layer number is fixed to be 7. It is because a larger number of layers introduce more parameters, which may lead to overfitting, while a small number of layers usually have inadequate parameters, which may result in underfitting.

It is usually difficult for the deep neural networks to afford insights toward interpreting mechanism since they introduce the complexity of interactions and nonlinearities.[30] Thus, it is also known as "black boxes" for a long time. To reduce the dimensionality of the data for better understanding how the CNN makes the decision, neighborhood components analysis (NCA) was employed to analyze the experimental XRD spectra.[31] As shown in the NCA map (Fig. 4), these XRD spectra from the same MOFs are clustered into ten separate groups, while the XRD spectra from different MOFs but with very similar spectra characteristics may result in overlapped or very close groups in the component map. It indicates that characteristics of main peaks such as position, intensity, and full width at half maximum—usually the main criteria to distinguish the XRD spectra—are reduced to show distinguished features as shown in the LLE component map.

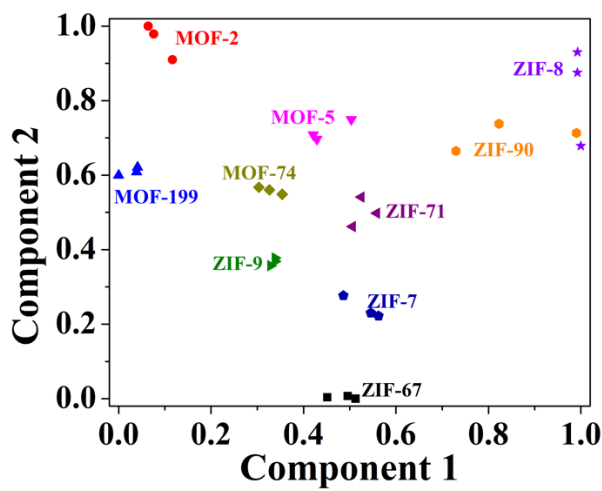

Figure 4. Neighborhood components analysis (NCA) map for clustering of XRD spectra of all MOFs samples.

In order to understand the mechanism of how the CNN model distinguishes individual experimental spectra, their CAMs in the fourteenth layers and corresponding XRD spectra were compared and shown in Fig. 5 and Fig. S3. In an image classification task, CAMs are usually used to reflect the main discriminative features of images, which helps interpret and improve classification accuracy.[2, 32] In our case, CAMs may afford a clear and direct impression on the characteristics of the XRD spectra that were the most relevant to the specific class. It is found that the red regions in the CAMs correspond to the main peaks of the XRD spectra. Hence, it is deduced that the CNN model can well distinguish XRD spectra according to the main peaks, i.e., the most important peaks. Further observation shows that MOF-74 (Fig. 5a) only exhibits six colorful regions including two dominated red regions, and ZIF-8 (Fig. 5b) exhibited much more colorful regions. This is similar to the way that a professional material scientist analyzes the spectra data. A traditional way to identify the compounds of interest and to map phase diagrams of combinatorial materials is to match descriptors of their XRD patterns such as peak positions, intensities, FWHM with a known database. Thus, it is straightforward to train machine learning models *via* data augmentation by peak scaling, peak elimination, and peak shift.[2, 16] The models successfully identify several crystal systems into which target materials are grouped. However, it is much more challenging to reach the goal of one-by-one classification, i.e., assign a correct label to individual XRD spectrum instead of seven crystal systems or 230 space groups among thousands of datasets. In our case, the trained CNN model can successfully and robustly perform one-by-one classification even though the experimental XRD spectra exhibit peak shift, scaling in peak intensities, or FWHM broadening compared to the theoretic spectra. As evident in Fig. 5 and Fig. S3, all XRD spectra of the MOFs exhibit peak shift, the existence of noise and peak intensity scaling compared to their theoretical spectra. Our model can tackle these abnormal phenomena and reach 96.7 % one-by-one classification accuracy. In addition, amazingly, the CNN can still identify the spectra even though when intensities of main peaks are largely changed. Take MOF-199 (Fig. 5c) and MOF-5 (Fig. 5d) as examples. The ratios of (220) to (222) peaks for experimental spectra of three MOF-199 samples are varied from 0.82 to 0.45. The trained model can all identify them in Top 3 (Table S4). The CAM on MOF-199 agrees well with this result, which shows that the dominated red region shifts to (222) from (220). Similarly, the ratios of (200) to (220) for MOF-5 are 0.29, 2.40, and 0.99 for M1, M2 and M3 of MOF-5, respectively. The model also can rank MOF-5 in Top 3 (Table S4). For MOF-5 M2 sample, the peak corresponding to the (400) plane increases to be the second highest peak. This result agrees well with the CAM analysis result and it shows that the two dominated regions correspond to the (200) and (400) peaks. In addition, we also observed that the lower crystallinity and existence of noise greatly affect the classification accuracy. Here, ZIF-9 was chosen as an example (Fig. S2f). The CNN model can classify the ZIF-9 M2 and ZIF-9 M3 in Top 2 and Top 4 rankings, but cannot distinguish ZIF-9 M1. The CAM analysis shows only one red dominated region, corresponding to the first highest peak. Due to the lower intensity and the existence of noise, the CNN could not learn the main features which can effectively assign all three samples to ZIF-9.

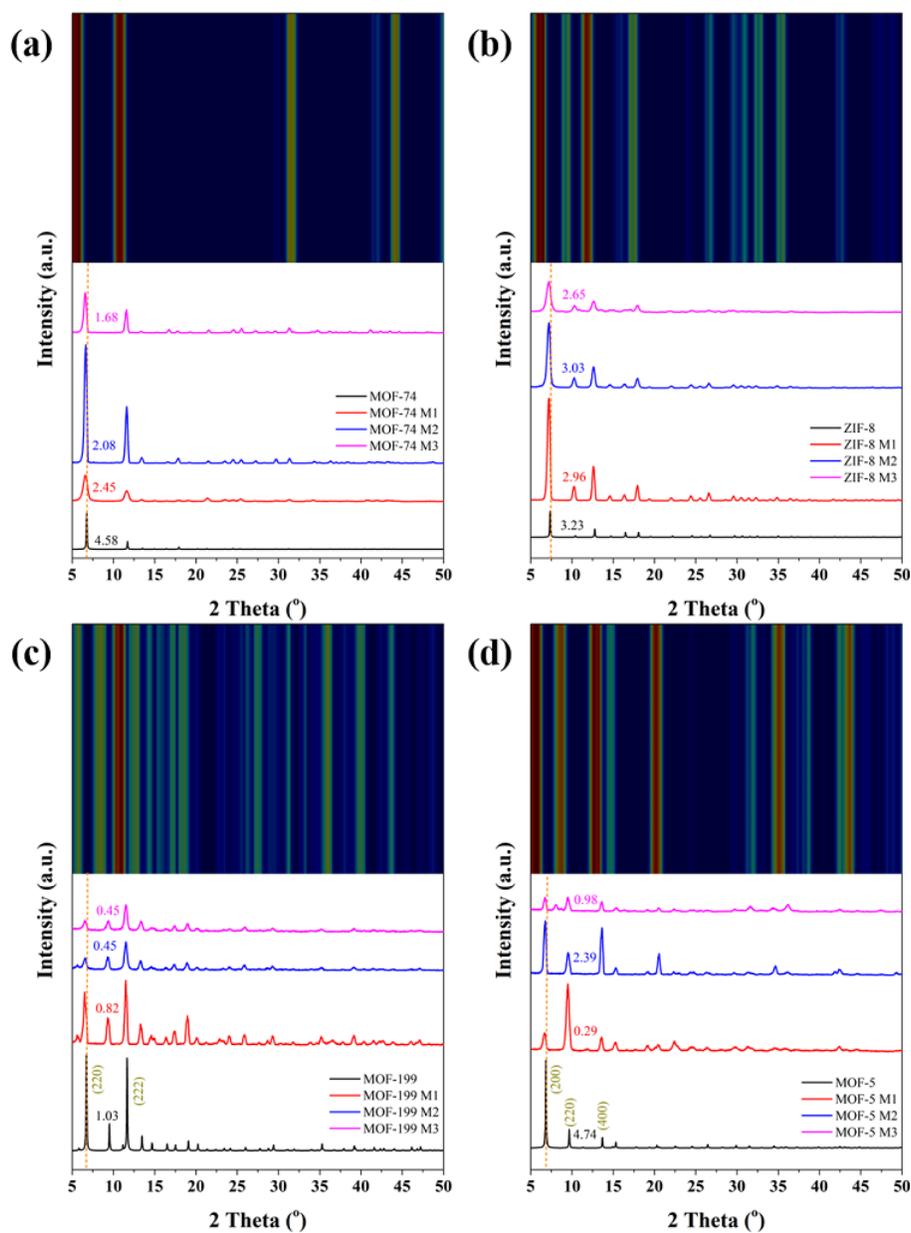

Figure 5. CAMs of the 14[th] layer output from the CNN model and corresponding XRD spectra: (a) MOF-74; (b) ZIF-8; (c) MOF-199; (d) MOF-5.

**Conclusion**

In summary, we demonstrated a CNN model trained by the theoretical XRD spectra augmented by noise from limited experimental spectra for rapid one-to-one identification of individual MOFs. CNN model is employed to identify XRD spectrum instead of categorizing them into groups or crystallinity systems. The optimized CNN model showed the highest identification accuracy of 96.7% for the Top 5 rankings among a dataset of 1012 XRD spectra. The advantages of the proposed CNN model can be summarized as follows: 1) it is a one-by-one identification instead of predicting several crystal groups; 2) the model was trained based on very limited theoretical data; 3) simple and straightforward noise-based data augmentation—not like the past technique that employed multi-step operations (peak scaling, elimination and shifting)—was deployed, thus it is easy to operate and requires less hyperparameter tuning; 4) the procedure of noise filtering can greatly increase the classification accuracy of CNN model. Finally, the proposed CNN model has the potential in not only numerous applications of XRD in materials science, but also the possible expansion of the solution to several other characterization techniques such as Raman, NMR, and FTIR.

**Methods**

**Chemicals.** $Zn(NO_3)_2 \cdot 6H_2O$ (Sigma Aldrich), $ZnCl_2 \cdot 6H_2O$ (Fluka), $Zn(CH_3COO)_2 \cdot 2H_2O$ (Fisher), $Co(NO_3)_2 \cdot 6H_2O$ (Sigma Aldrich), $CoCl_2 \cdot 6H_2O$ (Sigma Aldrich), $Co(CH_3COO)_2 \cdot 4H_2O$ (Fisher), 2-methylimidazole (Fisher), benzimidazole (Fisher), 4,5-dicholorimidazole (Fisher), 2-imidazolecarboxaldehyde (Fisher), terephthalic acid (BDC, Fisher), 1,3,5-benzentricarboxylic acid (BTC, Fisher), cetyltrimethylammonium bromide (CTAB, Sigma Aldrich), diethylamine (DEA, Fisher), triethylamine (TEA, Fisher), polyvinylpyrrolidone (PVP, $M_w$ = 360000 or 40000, Sigma Aldrich), N, N-dimethylformamide (DMF, Fisher), methanol (Fisher), and ethanol (Fisher) were used without any further purification.

**MOFs synthesis.** Here, all MOFs were synthesized by three different methods according to reported literatures. The detailed synthesis methods were described in Supporting Information.

**Characterization**. Powder X-ray diffraction (XRD) were obtained on a Bruker D8 Discover diffractometer (Cu Kα, λ=0.15406 nm).

**Machine learning models.**

Five classical machine learning algorithms, i.e., Naïve Bayes (NB), k-Nearest Neighbors (KNN), Logistic Regression (LOG), Random Forest (RF), and Support Vector Machine (SVM), were well-trained and employed for identifying XRD spectra. Feed-forward convolutional neural networks were constructed using Keras software library with the TensorFlow backend.[33] The machine learning models were performed using Python with Scikit-Learn on a high-performance computer with Intel i7-9700k CPU and Nvidia EVGA GeForce RTX 2070 GPU.

**Data availability**

All original XRD spectrum including theoretical and experimental data and Python scripts for preprocessing, augmentation and classification are available at https://github.com/javenlin/MOFs.

**Acknowledgement**

This work was supported by grants from US Department of Energy (Award number: DE-FE0031645) with Program Manager Karol K. Schrems, and National Science Foundation (Award number: 1825352) with Program Manager Khershed P. Cooper.

**Author contributions**

H. W proposed data augmentation method. He designed and trained CNN as well as implementing MLLE and activation mapping analysis. Y. X. synthesized and characterized MOFs. He also downloaded and processed the theoretic data. D. L. initially explored the idea by testing different types of machine learning models. Y. Z offered valuable suggestions in developing CNNs. M. X. provided valuable discussions and inspired solutions to the problems. J. L. conceived the idea, organized the research scopes, and oversaw all phases of the project. Y. X. wrote the manuscript which was edited by J. L. All authors commented on the manuscript.

**Competing interests**

The authors declare no competing interests.

**Additional information**

See supplemental information.